\documentstyle[epsfig,rotate]{aipproc}
\begin{document}
\title{
{\small\rm\vspace*{-3.2cm}
\rightline{TTP97-32}
\rightline{hep-ph/9708387}
\rightline{August 1997}
\ \\
\ \\}
Current Issues in CP Violation\footnote{Invited plenary talk given 
at {\it Beyond the Standard Model V}, 
Balholm, Norway, April 29 -- May 4, 1997. To appear in the proceedings.}
}

\author{Robert Fleischer\,\thanks{Address after September 1, 1997: Theory
Division, CERN, CH-1211 Geneva 23, Switzerland}}
\address{Institut f\"ur Theoretische Teilchenphysik\\
Universit\"at Karlsruhe\\
D-76128 Karlsruhe, Germany}

%\lefthead{LEFT head}
%\righthead{RIGHT head}
\maketitle

\begin{abstract}
A brief review of CP violation in $K$ and $B$ decays is given. While the 
observables $\varepsilon$ and Re$(\varepsilon'/\varepsilon)$ describing CP 
violation in neutral $K$ decays do not allow a powerful test of the CKM 
mechanism of CP violation, rare $K\to\pi\nu\overline{\nu}$ decays and in 
particular the $B$-meson system are much more promising in this respect. 
After a brief look at the $K$ system, selected aspects of CP-violating 
effects in $B$ decays are discussed. It is pointed out that combined 
branching ratios for $B\to\pi K$ modes, which have been observed recently 
by the CLEO collaboration, may allow to derive stringent constraints on the 
CKM angle $\gamma$ that could open a window to new physics.
\end{abstract}

\section*{Introduction}
The violation of the CP symmetry plays a central and fundamental role in 
modern particle physics. One of the reasons is that this phenomenon could 
guide us to physics beyond the Standard Model. To this end it is crucial 
to search for processes or relations among CP-violating observables that 
can be predicted in a clean way in the Standard Model framework and are 
not affected by hadronic uncertainties. 

Within the Standard Model of electroweak interactions \cite{sm}, CP violation 
is closely related to the Cabibbo-Kobayashi-Maskawa matrix (CKM matrix) 
\cite{cab,km} connecting the 
electroweak eigenstates of the $d$-, $s$- and $b$-quarks with their 
mass eigenstates. Whereas a single real parameter -- the Cabibbo angle -- 
is sufficient to parametrize the CKM matrix in the case of two fermion
generations \cite{cab}, three generalized Cabibbo-type angles and a single
{\it complex phase} are needed in the three generation case \cite{km}. This
complex phase is the origin of CP violation within the Standard Model. 

It turns out that CP-violating observables are proportional to the following
combination of CKM matrix elements: 
\begin{equation}
J_{\rm CP}=\pm\,\mbox{Im}\left(V_{i\alpha}V_{j\beta}V_{i\beta}^\ast 
V_{j\alpha}^\ast\right)\quad(i\not=j,\,\alpha\not=\beta)\,,
\end{equation}
which represents a measure of the ``strength'' of CP violation within the 
Standard Model \cite{jarlskog}. Since $J_{\rm CP}={\cal O}(10^{-5})$, CP 
violation is a small effect. In scenarios of new physics \cite{new-phys}, 
typically several new complex couplings are present yielding additional 
sources of CP violation.

\begin{figure}
\centerline{
\rotate[r]{
\epsfysize=6.5truecm
\epsffile{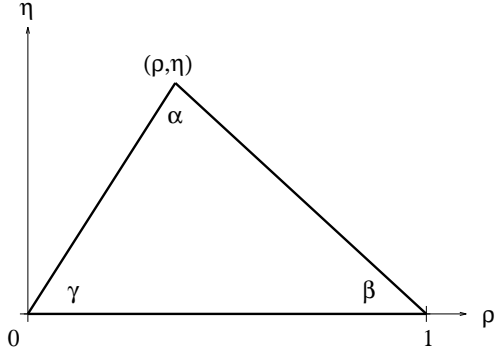}}}
\vspace{10pt}
\caption{The unitarity triangle of the CKM matrix in the $(\rho,\eta)$ plane.}
\label{UT:fig1}
\end{figure}

Concerning phenomenological applications, the parametrization 
\begin{equation}\label{wolf2}
\hat V_{\mbox{{\scriptsize CKM}}} =\left(\begin{array}{ccc}
1-\frac{1}{2}\lambda^2 & \lambda & A\lambda^3 R_b\, e^{-i\gamma} \\
-\lambda & 1-\frac{1}{2}\lambda^2 & A\lambda^2\\
A\lambda^3R_t\,e^{-i\beta} & -A\lambda^2 & 1
\end{array}\right)+\,{\cal O}(\lambda^4)
\end{equation}
with $\lambda=0.22$ and
\begin{eqnarray}\label{RbRt}
A&\equiv&\frac{1}{\lambda^2}\left|V_{cb}\right|=0.81\pm0.06\\
R_b&\equiv&\frac{1}{\lambda}\left|\frac{V_{ub}}{V_{cb}}\right|=
\sqrt{\rho^2+\eta^2}=0.36\pm0.08\\
R_t&\equiv&\frac{1}{\lambda}\left|\frac{V_{td}}{V_{cb}}\right|=
\sqrt{(1-\rho)^2+\eta^2}={\cal O}(1)
\end{eqnarray}
turns out to be very useful. This parametrization is a modification of the 
Wolfenstein parametrization \cite{wolf} making not only the hierarchy of the 
CKM elements, but also the dependence on the angles $\beta=\beta(\rho,\eta)$ 
and $\gamma=\gamma(\rho,\eta)$ of the usual ``non-squashed'' unitarity 
triangle of the CKM matrix explicit \cite{ut}. This triangle is a graphical 
illustration of the fact that the CKM matrix is unitary and is sketched 
for completeness in Fig.~\ref{UT:fig1}. 

\section*{A Brief Look at CP Violation in K Decays}
Although CP violation was discovered already in 1964 by observing $K_{\rm L}
\to\pi\pi$ decays \cite{ccft}, so far CP violation has been measured directly 
only within the neutral $K$-meson system, where it is described by two 
complex quantities called $\varepsilon$ and $\varepsilon'$ which are defined 
by the following ratios of decay amplitudes:
\begin{equation}\label{defs-eps}
\frac{A(K_{\rm L}\to\pi^+\pi^-)}{A(K_{\rm S}
\to\pi^+\pi^-)}=\varepsilon+\varepsilon',\quad
\frac{A(K_{\rm L}\to\pi^0\pi^0)}{A(K_{\rm S}
\to\pi^0\pi^0)}=\varepsilon-2\varepsilon'.
\end{equation}
While $\varepsilon=(2.26\pm0.02)\cdot e^{i\frac{\pi}{4}}\cdot10^{-3}$
parametrizes ``indirect'' CP violation originating from the fact that
the mass eigenstates of the neutral $K$-meson system are not eigenstates
of the CP operator, the quantity Re$(\varepsilon'/\varepsilon)$ provides
a measure of ``direct'' CP violation in $K\to\pi\pi$ transitions. The 
CP-violating observable $\varepsilon$ plays an important role to constrain
the unitarity triangle \cite{bf-rev} and informs us in particular about a 
non-vanishing, positive value of $\eta$. 

Despite enormous efforts, the experimental situation concerning 
Re$(\varepsilon'/\varepsilon)$ is still unclear at present. Whereas the 
CERN experiment NA31 finds Re$(\varepsilon'/
\varepsilon)=(23\pm7)\cdot10^{-4}$ indicating already direct CP violation, 
the result Re$(\varepsilon'/\varepsilon)=(7.4\pm5.9)\cdot10^{-4}$ of the 
Fermilab experiment E731 provides no unambiguous evidence for a non-zero 
effect. In the near future this situation will hopefully be clarified by 
improved measurements \cite{nguyen}. From a theoretical point of view, 
analyses of Re$(\varepsilon'/\varepsilon)$ are very involved and suffer at
present from large hadronic uncertainties \cite{bf-rev}. Consequently that 
observable does not allow a powerful test of the Standard Model description 
of CP violation unless the hadronic matrix elements of the relevant operators
are under better control. The major goal of a possible future observation of 
Re$(\varepsilon'/\varepsilon)\not=0$ would probably be the unambiguous 
exclusion of ``superweak'' models of CP violation \cite{superweak}.

More promising in respect of testing the CP-violating sector of the 
Standard Model are the rare decays $K_{\rm L}\to\pi^0\nu\overline{\nu}$ and
$K^+\to\pi^+\nu\overline{\nu}$. These decays, in particular the first one,
are very clean from a theoretical point of view \cite{bb-nlo}. Using the
top-quark mass $m_t$ and the CKM element $|V_{cb}|$ as an additional input, 
the branching ratios of these decays allow a determination of the unitarity 
triangle. A detailed analysis shows that in particular $\sin(2\beta)$ can be 
extracted with respectable accuracy \cite{bb}. Thus, comparing the value of 
$\sin(2\beta)$ determined that way with the one extracted from CP violation
in the ``gold-plated'' mode $B_d\to J/\psi\, K_{\rm S}$ (see the following
section) one has a powerful tool to probe physics beyond the Standard Model.
Unfortunately the branching ratios for $K_{\rm L}\to\pi^0\nu\overline{\nu}$ 
and $K^+\to\pi^+\nu\overline{\nu}$ are of ${\cal O}(10^{-11})$ and ${\cal 
O}(10^{-10})$, respectively, within the Standard Model \cite{bf-rev} making
measurements of these modes, especially of the former one, very challenging. 
Nevertheless there are plans to search for these important decays at BNL, 
FNAL and KEK. 

At present the observed CP violation in the $K$ system can be described
successfully by the Standard Model. This feature is, however, not surprising
since so far only a single CP-violating observable, $\varepsilon$, has to 
be fitted. Consequently many different non-standard model descriptions of 
CP violation are imaginable \cite{new-phys}. From the brief discussion given
above it is obvious that the $K$-meson system by itself cannot provide the 
whole picture of CP violation. Therefore it is essential to study CP 
violation outside this system. In this respect the $B$ system appears to 
be most promising which is also reflected by the tremendous experimental 
efforts at future $B$ factory facilities \cite{hitlin}. Let me note that 
there are also other interesting systems to investigate CP violation and 
to search for physics beyond the Standard Model, e.g.\ the $D$-meson system 
where sizable mixing or CP-violating effects would signal new physics because 
of the tiny Standard Model ``background'' \cite{D-rev}. Unfortunately I 
cannot discuss these systems in more detail in this presentation and shall 
focus on $B$ decays in the subsequent section. 

\section*{CP Violation in B Decays}
As far as CP violation and strategies for extracting angles of the unitarity 
triangle are concerned, the major role in the $B$ system is played by 
non-leptonic decays which can be divided into three decay classes: decays 
receiving both tree and penguin contributions, pure tree decays, and pure
penguin decays. There are two types of penguin topologies: gluonic (QCD)
and electroweak (EW) penguins originating from strong and electroweak 
interactions, respectively. Interestingly also the latter operators play 
an important role in several processes because of the large top-quark 
mass \cite{rev}.

In order to analyze non-leptonic $B$ decays theoretically, one uses low 
energy effective Hamiltonians that are calculated by making use of the 
operator product expansion yielding transition matrix elements of the
structure
\begin{equation}\label{ee2}
\langle f|{\cal H}_{\rm eff}|i\rangle\propto\sum_k C_k(\mu)
\langle f|Q_k(\mu)|i\rangle\,.
\end{equation}
The operator product expansion allows one to separate the short-distance
contributions to Eq.\ (\ref{ee2}) from the long-distance contributions
described by perturbative Wilson coefficient functions $C_k(\mu)$
and non-perturbative hadronic matrix elements $\langle f|Q_k(\mu)|
i\rangle$, respectively. As usual, $\mu$ denotes an appropriate
renormalization scale. Examples for such Hamiltonians and a  
discussion of the technicalities arising in calculations of Wilson
coefficients beyond the leading logarithmic approximation can be found in
a recent review \cite{bbl-rev}.

\subsection*{CP Asymmetries in Neutral B Decays}
A particular simple and interesting situation arises if we restrict ourselves
to decays of neutral $B_q$ mesons ($q\in\{d,s\}$) into CP self-conjugate 
final states $|f\rangle$ satisfying the relation $({\cal CP})|f\rangle=
\pm|f\rangle$. In that case the corresponding time-dependent CP asymmetry 
can be expressed as
\begin{eqnarray}
\lefteqn{a_{\mbox{{\scriptsize CP}}}(t)\equiv\frac{\Gamma(B^0_q(t)\to f)-
\Gamma(\overline{B^0_q}(t)\to f)}{\Gamma(B^0_q(t)\to f)+
\Gamma(\overline{B^0_q}(t)\to f)}=}\nonumber\\
&&{\cal A}^{\mbox{{\scriptsize dir}}}_{\mbox{{\scriptsize CP}}}(B_q\to f)
\cos(\Delta M_q\,t)+{\cal A}^{\mbox{{\scriptsize
mix--ind}}}_{\mbox{{\scriptsize CP}}}(B_q\to f)\sin(\Delta M_q\,t)
\,,\label{ee6}
\end{eqnarray}
where the direct CP-violating contributions have been separated from
the mixing-induced CP-violating contributions which are characterized by
\begin{equation}\label{ee7}
{\cal A}^{\mbox{{\scriptsize dir}}}_{\mbox{{\scriptsize CP}}}(B_q\to f)\equiv
\frac{1-\left|\xi_f^{(q)}\right|^2}{1+\left|\xi_f^{(q)}\right|^2}\quad
\mbox{and}\quad
{\cal A}^{\mbox{{\scriptsize mix--ind}}}_{\mbox{{\scriptsize
CP}}}(B_q\to f)\equiv\frac{2\,\mbox{Im}\,\xi^{(q)}_f}{1+\left|\xi^{(q)}_f
\right|^2}\,,
\end{equation}
respectively. Here direct CP violation refers to CP-violating effects
arising directly in the corresponding decay amplitudes, whereas mixing-induced
CP violation is related to interference between 
$B_q^0$--$\overline{B_q^0}$ mixing and decay processes. Note that the 
expression Eq.~(\ref{ee6}) has to be modified in the $B_s$ case 
for $t\gtrsim1/\Delta\Gamma_s$ because of the expected sizable width 
difference $\Delta\Gamma_s$ \cite{dun}. 

In general the observable
\begin{equation}
\xi_f^{(q)}\equiv e^{-i\phi_{\mbox{{\scriptsize M}}}^{(q)}}
\frac{A(\overline{B^0_q}\to f)}{A(B_q\to f)}\,,
\end{equation}
where 
\begin{equation}
\phi_{\mbox{{\scriptsize M}}}^{(q)}=\left\{\begin{array}{cr}
2\beta&\mbox{for $q=d$}\\
0&\mbox{for $q=s$}\end{array}\right.
\end{equation}
denotes the weak $B_q^0$--$\overline{B_q^0}$ mixing 
phase, suffers from large hardonic uncertainties that are introduced through 
the decay amplitudes $A$. There is, however, a very important special case 
where these uncertainties cancel. It is given if $B_q\to f$ is dominated by 
a single CKM amplitude. In that case $\xi_f^{(q)}$ takes the simple form
\begin{equation}\label{ee10}
\xi_f^{(q)}=\mp\exp\left[-i\left(\phi_{\mbox{{\scriptsize M}}}^{(q)}-
\phi_{\mbox{{\scriptsize D}}}^{(f)}\right)
\right],
\end{equation}
where $\phi_{\mbox{{\scriptsize D}}}^{(f)}$ is a characteristic weak decay
phase that is given by
\begin{equation}\label{e11}
\phi_{\mbox{{\scriptsize D}}}^{(f)}=\left\{\begin{array}{cc}
-2\gamma&\mbox{for dominant $\bar b\to\bar u\,u\,\bar r$ CKM amplitudes
in $B_q\to f$}\\
0&\,\mbox{for dominant $\bar b\to\bar c\,c\,\bar r\,$ CKM amplitudes
in $B_q\to f$.}
\end{array}\right.
\end{equation}
Here the label $r\in\{d,s\}$ distinguishes between $b\to d$ and $b\to s$
transitions. 

The most important application of this formalism is the ``gold-plated''
decay $B_d\to J/\psi\,K_{\rm S}$ \cite{csbs}. If one goes through the 
relevant Feynman diagrams contributing to this channel one finds that 
it is dominated to excellent accuracy by the $\bar b\to\bar cc\bar s$ 
CKM amplitude since penguins enter essentially with the same weak phase
as the leading tree contribution (see e.g.\ \cite{rev} for a recent 
detailed discussion). Therefore the weak decay phase vanishes and we get
\begin{equation}\label{ee12}
{\cal A}^{\mbox{{\scriptsize mix--ind}}}_{\mbox{{\scriptsize
CP}}}(B_d\to J/\psi\, K_{\mbox{{\scriptsize S}}})=+\sin[-(2\beta-0)]\,.
\end{equation}
This CP-violating observable allows hence a clean determination of the angle 
$\beta$ of the unitarity triangle (up to discrete ambiguities; see \cite{gq}
for a recent discussion) and is also very promising from an experimental 
point of view for future $B$ factories \cite{hitlin}.

Another important decay is $B_d\to\pi^+\pi^-$ which would measure 
$-\sin(2\alpha)$ in a clean way through
\begin{equation}\label{ee13}
{\cal A}^{\mbox{{\scriptsize mix--ind}}}_{\mbox{{\scriptsize
CP}}}(B_d\to\pi^+\pi^-)=-\sin[-(2\beta+2\gamma)]=-\sin(2\alpha)
\end{equation}
if there were no penguins present. However, penguins do contribute to 
$B_d\to\pi^+\pi^-$. The corresponding hadronic uncertainties
affecting the determination of $\alpha$ were discussed by many authors in 
the previous literature. There are even methods to control these 
uncertainties in a quantitative way. Unfortunately these strategies are 
usually rather challenging in practice. The most important examples 
are the $B\to\pi\pi$ isospin triangles proposed by Gronau and 
London~\cite{gl}, and the approach using $B\to\rho\,\pi$ modes suggested 
by Snyder and Quinn~\cite{sq}. An approximate method to correct for the 
penguin uncertainties in $B_d\to\pi^+\pi^-$ that appears to be promising 
for the early days of the $B$ factory era was proposed in \cite{fm1}. 
For a detailed discussion of these and other strategies the reader is 
referred to reviews on this subject, e.g.\ \cite{rev,cp-revs}. 

A decay appearing frequently as a tool to determine the angle $\gamma$ 
of the unitarity triangle is $B_s\to\rho^0 K_{\rm S}$. There, however, 
penguins are expected to lead to serious problems -- more serious than in
$B_d\to\pi^+\pi^-$ -- so that this mode appears to be the ``wrong'' way to 
extract $\gamma$ \cite{rev}. Other strategies allowing meaningful 
determinations of this angle will be discussed in a moment. 

\subsection*{CP Violation in Non-leptonic Penguin Modes}
In view of testing the Standard Model description of CP violation, 
penguin-induced modes play an important role. Because of the loop-suppression
of these ``rare'' FCNC processes, it is plausible -- and indeed the case in
specific model calculations -- that new physics contributions to these decays
are of similar magnitude as those of the Standard Model \cite{new-phys}. An 
example is the $b\to d$ penguin mode $B_d\to K^0\overline{K^0}$ (see e.g.\
\cite{mpw} for an analysis of new physics effects). If one assumes that 
penguins with internal top-quarks play the dominant role in this 
decay, the weak $B^0_d-\overline{B^0_d}$ mixing and $B_d\to K^0
\overline{K^0}$ decay phases cancel in the corresponding observable 
$\xi^{(d)}_{K^0\overline{K^0}}$ implying {\it vanishing} CP violation in 
that decay. Consequently one would conclude that a measurement of 
non-vanishing CP violation in $B_d\to K^0\overline{K^0}$
would signal physics beyond the Standard Model. However, long-distance 
effects related to penguins with internal charm- and up-quarks may easily
spoil the assumption of top-quark dominance \cite{rev,bf1}. These 
contributions may lead to sizable CP violation in $B_d\to K^0\overline{K^0}$ 
even within the Standard Model \cite{my-KK.bar}, so that a measurement of 
such CP asymmetries would not necessarily imply new physics as claimed in 
several previous papers. Unfortunately a measurement of these effects will 
be very difficult since the Standard Model expectation for the corresponding 
branching ratio is ${\cal O}(10^{-6})$ which is still one order of magnitude 
below the recent CLEO bound $\mbox{BR}(B_d\to K^0\overline{K^0})<1.7
\cdot10^{-5}$ \cite{cleo}. 

More promising in this respect and -- more importantly -- to search for 
physics beyond the Standard Model is the $b\to s$ penguin mode $B_d\to\phi\, 
K_{\rm S}$. The branching ratio for this decay is expected to be of 
${\cal O}(10^{-5})$ and may be large enough to investigate this channel 
at future $B$ factories. Interestingly there is to very good 
approximation no non-trivial CKM phase present in the corresponding decay 
amplitude \cite{rev}, so that direct CP violation vanishes and mixing-induced 
CP violation measures simply the weak $B^0_d-\overline{B^0_d}$ mixing phase 
which is related to the angle $\beta$ of the unitarity triangle. It should be 
stressed that this statement does {\it not} require the questionable 
assumption of top-quark dominance in penguin amplitudes. Consequently an 
important probe for new physics in $b\to s$ FCNC processes is provided by 
the relation
\begin{equation}
{\cal A}^{\mbox{{\scriptsize mix--ind}}}_{\mbox{{\scriptsize
CP}}}(B_d\to J/\psi\, K_{\mbox{{\scriptsize S}}})={\cal 
A}^{\mbox{{\scriptsize mix--ind}}}_{\mbox{{\scriptsize
CP}}}(B_d\to \phi\, K_{\mbox{{\scriptsize S}}})=-\sin(2\beta)\,,
\end{equation}
which holds within the Standard Model framework. The theoretical accuracy 
of this relation is limited by certain neglected terms that are 
CKM-suppressed by ${\cal O}(\lambda^2)$ and may lead to tiny direct 
CP-violating asymmetries in $B_d\to\phi\, K_{\rm S}$ of at most 
${\cal O}(1\%)$ \cite{rev}. Recently the importance of $B_d\to\phi\, 
K_{\rm S}$ and similar modes like e.g.\ $B_d\to\eta' K_{\rm S}$ to search 
for new physics in $b\to s$ transitions has been emphasized by several 
authors \cite{rev,BdPhiKs}.

\subsection*{A Closer Look at the $B_s$ System}
In the $B_s$ system very rapid $B^0_s-\overline{B^0_s}$ oscillations are
expected requiring an excellent vertex resolution system. Therefore studies
of CP violation in $B_s$ decays are regarded as being very difficult. An 
alternative route to investigate CP-violating effects may be provided by 
the width difference $\Delta\Gamma_s/\Gamma_s={\cal O}(20\%)$ arising from
CKM favoured $b\to c\bar c s$ transitions into final states that are
common both to $B^0_s$ and $\overline{B^0_s}$. Because of this width 
difference already {\it untagged} data samples of $B_s$ decays may exhibit 
CP-violating effects \cite{dun}. 

Recently several ``untagged'' strategies
to extract the CKM angle $\gamma$ were proposed using e.g.\ 
$B_s\to K^+K^-,K^0\overline{K^0}$ and $SU(3)$ flavor symmetry, angular
distributions in $B_s\to K^{\ast+}K^{\ast-},K^{\ast0}
\overline{K^{\ast0}}$ and $SU(2)$ isospin symmetry, or angular distributions 
in $B_s\to D^{\ast}\phi,\,D_s^{\ast\pm}K^{\ast\mp}$ allowing a clean 
determination of $\gamma$ \cite{fd}. Compared to the tagged case, such 
untagged measurements are obviously much more promising in view of efficiency, 
acceptance and purity. A lot of statistics is required, however,
and the natural place for these experiments seems to be a hadron machine.
The feasibility of untagged strategies to extract CKM phases depends 
crucially on a sizable width difference $\Delta\Gamma_s$ and it is not yet
clear whether it will turn out to be large enough to make these studies 
possible. 

The $B_s$ system provides also an important probe for physics beyond the
Standard Model through the decays $B_s\to D_s^{\ast+}D_s^{\ast-}$ and
$B_s\to J/\psi\,\phi$, which is the counterpart of the ``gold-plated'' mode
$B_d\to J/\psi\,K_{\rm S}$ to measure $\beta$. These modes are dominated by
a single CKM amplitude and allow -- in principal even from their untagged
data samples \cite{fd} -- the extraction of a CP-violating weak phase 
$\phi_{\rm CKM}\equiv2\lambda^2\eta$ which is expected to be of 
${\cal O}(0.03)$, i.e.\ very small, within the Standard Model \cite{bf-rev}. 
Consequently an extracted value of $\phi_{\rm CKM}$ that is much larger than 
this Standard Model expectation would signal new physics. Similarly as in 
the case of the penguin modes discussed above, it is plausible that physics 
beyond the Standard Model could also play an important role in the 
loop-suppressed $B^0_q-\overline{B^0_q}$ mixing processes \cite{new-phys}. 
In the case of the $B_s$ system, a sizable mixing phase could originate 
from new physics leading e.g.\ to significant CP violation in $B_s\to 
J/\psi\,\phi$. 

\subsection*{Extracting CKM Angles with Amplitude Relations}
Since mixing effects are absent in the charged $B$ system, the measurement of
a non-vanishing CP asymmetry in a charged $B$ decay would give unambiguous
evidence for direct CP violation thereby ruling out ``superweak'' models.
Such CP asymmtries arise from interference between decay amplitudes with
both different CP-violating weak and CP-conserving strong phases. Whereas
the weak phases are related to the CKM matrix, the strong phases are 
induced by strong final state interaction effects and introduce severe
theoretical uncertainties into the calculation destroying in general the
clean relation of the CP asymmetry to the phases of the CKM matrix. 

Nevertheless there are decays of charged $B$ mesons which play an important
role to extract angles of the unitarity triangle, in particular for $\gamma$.
To this end amplitude relations -- either exact or approximate ones 
based on flavor symmetries -- are used. A recent review of these methods
can be found in \cite{rev}. The ``prototype'' is the approach to determine 
$\gamma$ with the help of triangle relations among $B^\pm\to D K^\pm$ decay 
amplitudes proposed by Gronau and Wyler \cite{gw}. Unfortunately the 
corresponding triangles are expected to be very ``squashed'' ones. Moreover
one has to deal with additional experimental problems \cite{ads}, so that 
this approach is very difficult from a practical point of view. Recently 
more refined variants have been proposed by Atwood, Dunietz and Soni 
\cite{ads}.

About three years ago, several methods to extract CKM angles were presented
by Gronau, Hern\'andez, London and Rosner who have combined the $SU(3)$
flavor symmetry of strong interactions with certain dynamical assumptions
to derive relations among $B\to\pi\pi,\pi K,K\overline{K}$ decay 
amplitudes \cite{ghlr}. This approach has been very popular over recent years
and requires only a measurement of the relevant branching ratios. A closer 
look shows, however, that it suffers despite its attractiveness from several 
problems: the $SU(3)$ relations are not valid exactly, QCD penguins with 
internal charm- and up-quarks play in certain cases an important role, and 
interestingly also EW penguins lead to complications. In order to eliminate
the EW penguin contributions, usually very involved strategies are needed. 
A detailed discussion of all these methods (mainly to extract $\gamma$) is 
beyond the scope of this article and the reader is referred to a recent 
review \cite{rev} and references therein. 

\subsection*{Searching for $\gamma$ and New Physics with $B\to\pi K$ Modes}
A simple approach to determine $\gamma$ with the help of the branching ratios 
for $B^+\to\pi^+K^0$, $B^0_d\to\pi^-K^+$ and their charge conjugates was 
proposed in \cite{PAPIII} (see also \cite{rev}). It makes use of the fact 
that the general phase 
structure of the corresponding decay amplitudes is known reliably within the 
Standard Model, and employs the $SU(2)$ isospin symmetry of strong interactions
to relate the QCD penguin contributions. If the magnitude of the 
current-current amplitude $T'$ contributing to $B^0_d\to\pi^-K^+$ is known --
it can be fixed e.g.\ through $B^+\to\pi^+\pi^0$, ``factorization'', or 
hopefully lattice gauge theory one day -- two amplitude triangles can be 
constructed allowing in particular the extraction of $\gamma$. This approach 
is promising for future $B$ physics experiments, since it requires only 
time-independent measurements of branching ratios at the ${\cal O}(10^{-5})$ 
level. If one measures in addition the branching ratios for $B^+\to\pi^0K^+$ 
and its charge-conjugate, also the $b\to s$ EW penguin amplitude can be 
determined which is an interesting probe for new physics \cite{PAPI}.

\begin{figure}
\centerline{\epsfig{file=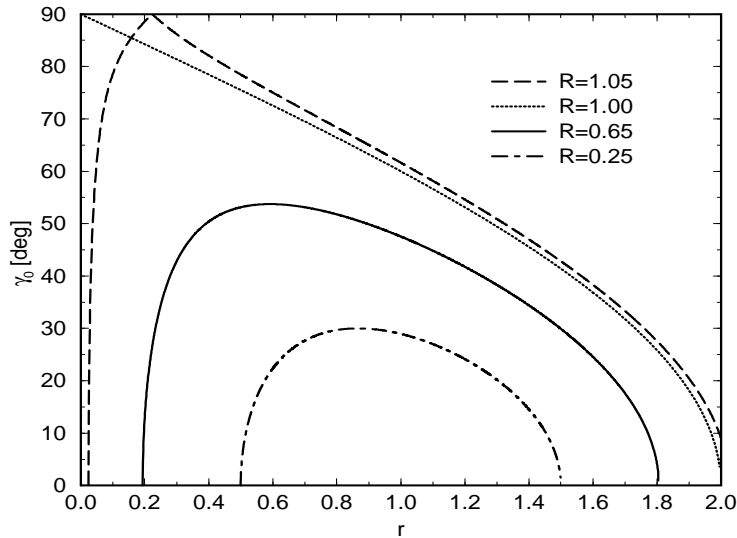,height=3.3in,width=4.5in}}
\vspace{10pt}
\caption{The dependence of $\gamma_0$ on the amplitude ratio $r$ for various
values of $R$. The range of $R$ corresponds to recent CLEO measurements.}
\label{gamma:fig2}
\end{figure}

Recently the CLEO collaboration has reported the first observation of the 
decays $B^+\to\pi^+K^0$ and $B^0_d\to\pi^-K^+$ \cite{cleo}. At present, 
however, only combined branching ratios, i.e.\ averaged ones over decays 
and their charge-conjugates, are available with large experimental 
uncertainties. Therefore it is unfortunately not yet possible to extract 
$\gamma$ from the triangle construction proposed in \cite{PAPIII}. The 
recent CLEO measurements allow, however, to derive interesting 
{\it constraints} on $\gamma$ which are of the form 
\begin{equation}\label{gamma-bound}
0^\circ\leq\gamma\leq\gamma_0\quad\lor\quad180^\circ-
\gamma_0\leq\gamma\leq180^\circ 
\end{equation}
and are hence complementary to the presently 
allowed range 
\begin{equation}\label{UT-fits}
42^\circ\lesssim\gamma\lesssim135^\circ
\end{equation}
for that angle arising from the usual fits of the unitarity triangle 
\cite{bf-rev}. This remarkable feature has been pointed out recently by
Mannel and myself \cite{fm2}. The quantity $\gamma_0$ in Eq.\ 
(\ref{gamma-bound}) depends both on the ratio 
\begin{equation}
R\equiv\frac{\mbox{BR}(B_d\to\pi^\mp K^\pm)}{\mbox{BR}(B^\pm\to\pi^\pm K)} 
=\frac{\mbox{BR}(B_d^0\to\pi^- K^+)+\mbox{BR}(\overline{B_d^0}\to\pi^+ 
K^-)}{\mbox{BR}(B^+\to\pi^+K^0)+\mbox{BR}(B^-\to\pi^-\overline{K^0})}
\end{equation}
of the combined branching ratios and on the amplitude ratio 
\begin{equation}
r\equiv|T'|/| P'|
\end{equation} 
of the current-current and penguin operator contributions to 
$B_d\to\pi^\mp K^\pm$ as can be seen in Fig.~\ref{gamma:fig2}. If we look 
at that figure we observe that $R=1$ is a very important special case.
For $R>1$ constraints on $\gamma$ require some knowledge about $r$, i.e.\
$|T'|$, e.g.\ from $B^+\to\pi^+\pi^0$, ``factorization'', or hopefully 
lattice gauge theory one day. On the other hand, if $R$ is found 
experimentally to be smaller than one, bounds on $\gamma$ can always be 
obtained independent of $r$. The point is that $\gamma_0$ 
takes a maximal value 
\begin{equation}
\gamma_0^{\rm max}=\mbox{arccos}(\sqrt{1-R})
\end{equation}
depending only on the ratio $R$ of combined $B\to\pi K$ branching ratios 
\cite{fm2}. 

Let us take as an example the central values of the recent CLEO
measurements \cite{cleo} yielding $R=0.65$. This value corresponds to 
$\gamma_0^{\rm max}=54^\circ$ and implies the range $0^\circ\leq\gamma\leq
54^\circ$ $\lor$ $126^\circ\leq\gamma\leq180^\circ$ which has only the small
overlap $42^\circ\lesssim\gamma\leq54^\circ$ $\lor$ $126^\circ\leq\gamma
\lesssim135^\circ$ with the range (\ref{UT-fits}). The two pieces of this
range are distinguished by the sign of a quantity $\cos\delta$, where 
$\delta$ is the
CP-conserving strong phase shift between the $T'$ and $P'$ amplitudes. Using
arguments based on ``factorization'' one expects $\cos\delta>0$ corresponding 
to the former interval of that range, i.e.\ $42^\circ\lesssim\gamma\leq
54^\circ$ in our example \cite{fm2} (see \cite{ag} for a recent model 
calculation). Consequently, once more data come in confirming $R<1$, 
the decays $B_d\to\pi^\mp K^\pm$ 
and $B^\pm\to\pi^\pm K$ may put the Standard Model to a decisive test and 
could open a window to new physics. The implications of physics beyond the 
Standard Model on the $B\to\pi K$ modes have been analyzed in a recent 
paper \cite{fm3}. A striking new physics effect would e.g.\ be sizable CP 
violation in the decay $B^+\to\pi^+K^0$.

Let me finally note that the consistent description of $B^\pm\to\pi^\pm K$ and
$B_d\to\pi^\mp K^\pm$ within the Standard Model implies in addition to the 
constraints on $\gamma$ discussed above the range
\begin{equation}
\left|1-\sqrt{R}\right|\leq r\leq1+\sqrt{R}
\end{equation}
for the amplitude ratio $r$ and upper limits for the CP-violating asymmetry 
in $B^0_d\to\pi^-K^+$. It is interesting to note that commonly accepted means 
to estimate $r$ yield values that are at the edge of compatibility with the 
present CLEO measurements \cite{fm2}. 

\vspace{0.2truecm}

In conclusion, I hope that the aspects of CP violation in $K$ and $B$ decays
that I have selected for this presentation have convinced the reader that 
this phenomenon provides powerful tools to probe new physics. More 
advanced experimental studies of CP-violating effects in the Kaon system 
and the exploration of CP violation at $B$ physics facilities are just ahead 
of us. In the foreseeable future these experiments may bring unexpected 
results that could shed light on physics beyond the Standard Model. Certainly 
the coming years will be very exciting!

\section*{Acknowledgments}
I am very grateful to Gerald Eigen and Per Osland for inviting me to that
most enjoyable conference and for providing generous travel support.

\end{document}